\documentstyle[12pt,amsfonts]{article}
\newcommand{\svec}[1]{ \stackrel{\rightarrow}{#1} }

\newcommand{\define}{ \stackrel{\triangle}{=} }

\def\be{\begin{equation}}
\def\ee{\end{equation}}
\def\ba{\begin{array}}
\def\ea{\end{array}}

\def\d4{{\rm d}^4}

\begin{document}
\title{\bf Non-Relativistic Limit of Dirac Equations in Gravitational
            Field and Quantum Effects of Gravity }
\author{{Ning WU}
\thanks{email address: wuning@mail.ihep.ac.cn}
\\
\\
{\small Institute of High Energy Physics, P.O.Box 918-1,
Beijing 100039, P.R.China}
\thanks{mailing address}}
\maketitle

~~


\begin{abstract}
Based on unified theory of electromagnetic interactions and
gravitational interactions, the non-relativistic limit of the
equation of motion of a charged Dirac particle in gravitational
field is studied. From the Schrodinger equation obtained from
this non-relativistic limit, we could  see that the classical
Newtonian gravitational potential appears as a part of the potential
in the Schrodinger equation, which can explain the gravitational
phase effects found in COW experiments. And because of this
Newtonian gravitational potential, a quantum particle in earth's
gravitational field may form a gravitationally bound quantized
state, which had already been detected in experiments. Three different
kinds of phase effects related to gravitational interactions
are studied in this paper, and these phase effects should
be observable in some astrophysical processes.
Besides, there exists direct coupling between gravitomagnetic
field and quantum spin, radiation caused by this coupling
can be used to directly determine the gravitomagnetic
field on the surface of a star.
\\

\end{abstract}

\newpage

\Roman{section}

\section{Introduction}

For a long time, experimental studies of gravity are mainly concentrated
on classical tests of gravity, which aimed to test classical Newton's
theory of gravity and Einstein's general relativity. Of course, both
classical Newton's theory of gravity and Einstein's general relativity
have met with brilliant series of successes in describing phenomena
related to classical gravitational interactions. \\

However, just study gravity at classical level is not enough for
us to understand the nature of  gravity. From the point of view of
quantum field theory, interactions are transmitted by some kinds
of fields. After quantization, such fields consist of
elementary particles. For example, electromagnetic field consists
of photons. Similarly, gravitational field should consist of
gravitons. In order to study gravity in quantum mechanics, a quantum
theory of gravity which is perturbatively renormalizable is necessary.
And in order to test the theory of gravity in quantum level, experimental
study on quantum effects of gravity is highly essentially important.
Without doubt, studying gravity in quantum level will push our further
understanding on the nature of gravity, and should be one of
the main stream of the future study on the nature of gravity
from the physical point of view. Up to now, there are also some
experimental studies on quantum effects of a quantum state in
gravitational field, such as gravitational phase effects found in
COW experiments\cite{1,2,3} and gravitationally bound quantized
states found recently\cite{4,5}. \\

Because of the tremendous success of the standard model in elementary
particle physics, it is now generally believed that the common nature
of the fundamental interactions should be gauge. Based on gauge principle,
quantum gauge theory of gravity is proposed to study quantum
behavior of gravity in the framework of quantum field
theory\cite{6,7,8,9,10,11}.  At classical level, quantum gauge
theory of gravity gives out Einstein's general relativity\cite{6,7},
and can explain classical tests of gravity. At quantum level,
it is a perturbatively renormalizable quantum theory of gravity,
and can be used to explain quantum behavior
of gravitational interactions. In contrast with the traditional
gauge gravity, quantum gauge theory of gravity is a physical theory
of gravity, not a geometric theory of gravity, in which, four kinds
of fundamental interactions in nature can be unified in a simple
and beautiful way\cite{12,13,14}. Massive gravitons can also be
used to explain possible origin of dark matter and dark energy\cite{15,16}.
Gravitational phase effects\cite{1,2,3} and gravitationally bound
quantized states\cite{4,5} can also be explained in a more natural
and fundamental way, which will be discussed in this paper.
Besides, there exists another interesting
quantum phenomenon of gravity, that is, after considering spontaneously
symmetry breaking, non-homogenous super currents may shield gravity\cite{17}.
Gravitational shielding effects may have already been found in
experiments\cite{18,19}.
\\

The close analogy between Newton's gravitation law and Coulomb's law
of electricity led to the birth of the concept of gravitomagnetism
in the nineteenth century\cite{m01,m02,m03,m04}. Later, in the
framework of General Relativity, gravitomagnetism was extensively
explored\cite{m05,m06,m07} and recently some experiments are
designed to test gravitomagnetism effects. Some recently reviews
on gravitomagnetism can be found in literatures \cite{m08,m09,m10}.
In quantum gauge theory of gravity, gravitoelectric field and
gravitomagnetic field are naturally defined as component of
field strength of gravitational gauge field.
\\

In this paper, non-relativistic limit of Dirac equations in gravitational field
is studied, which is the basis in quantum gauge theory of gravity
to explain gravitational phase effects found in COW experiments\cite{1,2,3}
and gravitationally bound quantized states found recently\cite{4,5}.
Besides, two new kinds of gravitational phase effects, including
gravitational Aharonov-Bohm effect, are also discussed.
The explanation given in this paper
is from first principle, not just explain it by phenomenologically
introduce classical Newtonian gravitational potential into Schrodinger
equation. \\

\section{Unified Theory of Electromagnetic Interactions and
        Gravitational Interactions}

First, for the sake of integrity, we give a simple introduction
to gravitational gauge theory and introduce some notations which
is used in this paper. Details on quantum gauge theory of gravity
can be found in literatures \cite{6,7,8,9,10,11}.
In gauge theory of gravity, the most fundamental
quantity is gravitational gauge field $C_{\mu}(x)$,
which is the gauge potential corresponding to gravitational
gauge symmetry. Gauge field $C_{\mu}(x)$ is a vector in
the corresponding Lie algebra, which
is called the gravitational Lie algebra.
So $C_{\mu}(x)$ can expanded as
\be \label{2.10}
C_{\mu}(x) = C_{\mu}^{\alpha} (x) \hat{P}_{\alpha},
~~~~~~(\mu, \alpha = 0,1,2,3)
\ee
where $C_{\mu}^{\alpha}(x)$ is the component field and
$\hat{P}_{\alpha} = -i \frac{\partial}{\partial x^{\alpha}}$
is the  generator of the global gravitational gauge group.
The gravitational gauge covariant derivative is defined by
\be \label{2.9}
D_{\mu} = \partial_{\mu} - i g C_{\mu} (x)
= G_{\mu}^{\alpha} \partial_{\alpha},
\ee
where $g$ is the gravitational coupling constant and matrix
$G$ is given by
\be \label{2.11}
G = (G_{\mu}^{\alpha}) = ( \delta_{\mu}^{\alpha} - g C_{\mu}^{\alpha} )
=(I -g C)^{\alpha}_{\mu}.
\ee
Matrix $G$ is an important quantity in gauge theory
of gravity. Its inverse matrix is denoted as $G^{-1}$
\be \label{2.12}
G^{-1} = \frac{1}{I - g C} = (G^{-1 \mu}_{\alpha}).
\ee
Using matrix $G$ and $G^{-1}$, we can define two important
quantities
\be \label{2.13}
g^{\alpha \beta} = \eta^{\mu \nu}
G^{\alpha}_{\mu} G^{\beta}_{\nu},
\ee
\be \label{2.14}
g_{\alpha \beta} = \eta_{\mu \nu}
G_{\alpha}^{-1 \mu} G_{\beta}^{-1 \nu}.
\ee
\\

The  field strength of gravitational gauge field is defined by
\be \label{2.16}
F_{\mu\nu} \define \frac{1}{-ig} \lbrack D_{\mu}~~,~~D_{\nu} \rbrack.
\ee
Its explicit expression is
\be \label{2.17}
F_{\mu\nu}(x) = \partial_{\mu} C_{\nu} (x)
-\partial_{\nu} C_{\mu} (x)
- i g C_{\mu} (x) C_{\nu}(x)
+ i g C_{\nu} (x) C_{\mu}(x).
\ee
$F_{\mu\nu}$ is also a vector in gravitational Lie algebra
and can be expanded as,
\be \label{2.18}
F_{\mu\nu} (x) = F_{\mu\nu}^{\alpha}(x) \cdot \hat{P}_{\alpha},
\ee
where
\be \label{2.19}
F_{\mu\nu}^{\alpha} = \partial_{\mu} C_{\nu}^{\alpha}
-\partial_{\nu} C_{\mu}^{\alpha}
-  g C_{\mu}^{\beta} \partial_{\beta} C_{\nu}^{\alpha}
+  g C_{\nu}^{\beta} \partial_{\beta} C_{\mu}^{\alpha}.
\ee
\\

The $U(1)$ gauge invariant field strength of electromagnetic field
is given by\cite{12,20}
\be \label{2.20}
{\mathbb A}_{\mu \nu} = A_{\mu \nu}
+ g G^{-1 \lambda}_{\alpha} A_{\lambda} F_{\mu\nu}^{\alpha},
\ee
where $A_{\mu\nu}$ is defined by
\be \label{2.21}
A_{\mu \nu} = (D_{\mu} A_{\nu}) - (D_{\nu} A_{\mu}).
\ee
The symmetry group of the unified theory of electromagnetic
interactions and gravitational interactions is the  semi-direct
product group of $U(1)$ Abelian  group and
gravitational gauge group. We denote it as
\be \label{2.22}
GU(1) \define
U(1) \otimes_s Gravitational~Gauge~Group = \lbrace g(x)\rbrace.
\ee
$GU(1)$ gauge covariant derivative
is
\be \label{2.23}
{\mathbb D}_{\mu} = \partial_{\mu}
- i e A_{\mu} - i g C_{\mu},
\ee
where $e$ is the electromagnetic coupling constant. For electron,
$-e$ should be used instead of $e$, for it carries negative
electric charge.
\\

In the unified theory of electromagnetic interactions and gravitational
interactions, after considering Dirac Field, the Lagrangian
is selected to be\cite{20,6,7}
\be \label{2.24}
{\cal L}_0 = - \bar{\psi}
( \gamma^{\mu}  {\mathbb D}_{\mu}  + m ) \psi
- \frac{1}{4} \eta^{\mu \rho} \eta^{\nu \sigma}
{\mathbb A}_{\mu \nu} {\mathbb A}_{\rho \sigma}
 - \frac{1}{4} C^{\mu\nu\rho\sigma}_{\alpha\beta}
F^{\alpha}_{\mu \nu} F^{\beta}_{\rho \sigma} ,
\ee
where
\be
C^{\mu\nu\rho\sigma}_{\alpha\beta}
= \frac{1}{4} \eta^{\mu \rho}
\eta^{\nu \sigma} g_{\alpha \beta}
+ \frac{1}{2} \eta^{\mu \rho}
G^{-1 \nu}_{\beta} G^{-1 \sigma}_{\alpha}
- \eta^{\mu \rho}
G^{-1 \nu}_{\alpha} G^{-1 \sigma}_{\beta}.
\label{2.25}
\ee
The action of the system is defined by
\be \label{2.26}
S = \int \d4 x  \sqrt{- {\rm det} g_{\alpha \beta} }~ {\cal L}_0.
\ee
It can be strictly proved that this action has strict local
gravitational gauge symmetry and $U(1)$ gauge symmetry.\\

From this action, we can deduce the Euler-Lagrangian equation
of motion of Dirac field, which is
\be \label{2.27}
\left\lbrack \gamma^{\mu} ( \partial_{\mu} - ie A_{\mu}
- g C_{\mu}^{\alpha} \partial_{\alpha} ) + m \right\rbrack
\psi = 0.
\ee
This equation of motion describes the electromagnetic interactions
and gravitational interactions of a Dirac particle. In this paper,
we will discuss its non-relativistic limit, which is the basis
of our understanding on many quantum  effects of gravity which
are detected by experiments. \\

\section{Non-Relativistic Limit of Dirac Equation}
\setcounter{equation}{0}

We start our discussion from eq.(\ref{2.27}), which can be written
into another form
\be \label{6.1}
\lbrack \gamma^{\mu} (D_{\mu} - i e A_{\mu})
+ m \rbrack \psi= 0.
\ee
Define
\be \label{6.2}
\alpha^i = - \gamma^0 \gamma^i,
\ee
then, we can change eq.(\ref{6.1}) into
\be \label{6.3}
\left \lbrack ( D_0 - ie A_0 ) + \alpha^i
(D_i - i e A_i) - \gamma^0  m \right\rbrack \psi = 0.
\ee
Multiply both sides of eq.(\ref{6.3}) with
\be \label{6.4}
\left \lbrack ( D_0 - ie A_0 ) - \alpha^i
(D_i - i e A_i) + \gamma^0  m \right\rbrack,
\ee
we will get
\be \label{6.5}
\ba{l}
\left\lbrace (D_0 - i e A_0 )^2 -
\left\lbrack \svec{\alpha} \cdot (\svec{D} - i e \svec{A})
\right\rbrack ^2 + m^2 \right.\\
\\
+ \left. (D_0 - i e A_0 ) \left\lbrack \svec{\alpha} \cdot
(\svec{D} - i e \svec{A}) \right\rbrack -
\left \lbrack \svec{\alpha} \cdot (\svec{D} - i e \svec{A})
\right \rbrack (D_0 - i e A_0 ) \right\rbrace \psi = 0.
\ea
\ee
Define\cite{7,21},
\be \label{6.6}
E_i = F_{0i} = E_i^{\alpha} \hat P_{\alpha},
\ee
\be \label{6.7}
E_{ei} = A_{i0},
\ee
\be \label{6.8}
B_i = - \frac{1}{2} \varepsilon_{ijk} F_{jk} =
 B_i^{\alpha} \hat P_{\alpha},
\ee
\be \label{6.9}
B_{ei} =  \frac{1}{2} \varepsilon_{ijk} A_{jk}.
\ee
$E_i^{\alpha}$ is the gravitoelectric field and $B_i^{\alpha}$ is the
gravitomagnetic field, while $E_{ei}$ is the electromagnetic
electric field and $B_{ei}$ is the electromagnetic magnetic field.
The $\alpha = 0$ components of gravitoelectromagnetic fields
$E_i^0$ and $B_i^0$ correspond to those discussed in
literatures \cite{m05,m06,m07,m08,m09,m10}.
Using the following relation
\be \label{6.10}
(\svec{\alpha} \cdot \svec{A})
(\svec{\alpha} \cdot \svec{B})
= (\svec{A} \cdot \svec{B})
+ 2 i \svec{\Sigma} \cdot
(\svec{A} \times \svec{B}),
\ee
where $\svec{\Sigma}$ is the spin matrix of Dirac field, we can get
\be \label{6.11}
\ba{rcl}
\left\lbrack \svec{\alpha} \cdot
(\svec{D} - i e \svec{A} ) \right\rbrack ^2
&= &(\svec{D} - i e \svec{A} )^2
+ 2 i \svec{\Sigma} \cdot
\left\lbrack (\svec{D} -  i e \svec{A} ) \times
(\svec{D} - i e \svec{A} ) \right\rbrack  \\
&&\\
&=& (\svec{D} - i e \svec{A} )^2
- 2 g \svec{\Sigma} \cdot \svec{B}
+ 2 e \svec{\Sigma} \cdot \svec{B}_e.
\ea
\ee
On the other hand, it is easy to prove that
\be \label{6.12}
( D_0 - i e A_0 ) \left\lbrack \svec{\alpha} \cdot
(\svec{D} - i e \svec{A}) \right\rbrack -
\left \lbrack \svec{\alpha} \cdot (\svec{D} - i e \svec{A})
\right \rbrack (D_0 - i e A_0 ) =
- i g \svec{\alpha} \cdot \svec{E}
+ i e \svec{\alpha} \cdot \svec{E}_e.
\ee
Then, Dirac equation eq.(\ref{6.5}) is changed into
\be \label{6.13aa}
\left\lbrack (D_0 - i e A_0 )^2 -
(\svec{D} - i e \svec{A})^2
+ 2 g \svec{\Sigma} \cdot \svec{B}
- 2 e \svec{\Sigma} \cdot \svec{B}_e
- i g \svec{\alpha} \cdot \svec{E}
+ i e \svec{\alpha} \cdot \svec{E}_e
+ m^2 \right\rbrack \psi = 0.
\ee
\\

Define
\be \label{6.13a}
\psi  = \left (
\ba{c}
\varphi \\
\chi
\ea
\right ),
\ee
where $\varphi$ is the wave function for the Dirac particle and
$\chi$ is for its antiparticle. Both of them are two-component
spinors.  Using
\be \label{6.13b}
\svec{\alpha} = \left (
\ba{cc}
- \svec{\sigma} & 0\\
0 & \svec{\sigma}
\ea
\right ),
\ee
\be \label{6.13c}
\svec{\Sigma} = \frac{1}{2}  \left (
\ba{cc}
 \svec{\sigma} & 0\\
0 & \svec{\sigma}
\ea
\right ),
\ee
we could obtain the following equation for the field $\varphi$
\be \label{6.13}
\left\lbrack (D_0 - i e A_0 )^2 -
(\svec{D} - i e \svec{A})^2
+ g \svec{\sigma} \cdot \svec{B}
- e \svec{\sigma} \cdot \svec{B}_e
+ i g \svec{\sigma} \cdot \svec{E}
- i e \svec{\sigma} \cdot \svec{E}_e
+ m^2 \right\rbrack \varphi = 0.
\ee
\\

Eq.(\ref{6.13}) is a strict relativistic equation. Now, let's
discuss non-relativistic limit of it. Suppose that the moving
speed of the Dirac particle is slow, then, in non-relativistic
limit, we can define,
\be \label{6.14}
\varphi (\svec{x},t) = \Psi (\svec{x},t) e^{-i mt},
\ee
where $\Psi (\svec{x},t)$ satisfies Schrodinger equation,
\be \label{6.15}
i \frac{\partial}{\partial t} \Psi (\svec{x},t)
= E \Psi (\svec{x},t) ,
\ee
with $E$ the kinematical energy of the Dirac particle, which is
much smaller than its mass $m$. Using the following equation
\be \label{6.16}
(D_0 - ie A_0)^2 \psi \cong
\left\lbrace \left\lbrack
- 2im \partial_0
- 2em A_0 - m^2 ( 1 - 2g C_0^0 )
\right\rbrack \Psi \right\rbrace e^{-imt},
\ee
\be \label{6.16a}
(\svec{D} - ie \svec{A})^2 \psi =
\left\lbrace \left\lbrack
\svec{D} -ie \svec{A} + i m g \svec{C}^0
\right\rbrack ^2 \Psi \right\rbrace e^{-imt},
\ee
we can change eq.(\ref{6.13}) into
\be \label{6.17}
\ba{rcl}
i \frac{\partial \Psi}{\partial t} &=& \left\lbrack
\frac{1}{2m} \left (-i \svec{D} - e \svec{A}
        + m g \svec{C}^0 \right)^2
+ m g C_0^0 - e A_0  \right.\\
&&\\
&& \left.
- \frac{g}{2} \svec{\sigma} \cdot \svec{B}^0
- \frac{e}{2 m} \svec{\sigma} \cdot \svec{B}_e
- \frac{i g}{2}  \svec{\sigma} \cdot \svec{E}^0
- \frac{i e }{ 2 m}\svec{\sigma} \cdot \svec{E}_e
 \right\rbrack \Psi,
\ea
\ee
where
\be
\svec{C}^0 = (C_i^0).
\ee
It could be seen that the classical Newtonian gravitational potential
naturally enters into the Schrodinger equation. Besides, there is direct
coupling between spin and gravitomagnetic field, no matter the
Dirac particle carries electric charge or not. \\

\section{Quantum  Effects of Gravity}
\setcounter{equation}{0}

In most cases, the gravitational field is very weak, for example,
on earth,
\be \label{4.1}
g C^0_0 = \frac{GM}{r} \sim 7 \times 10^{-10}.
\ee
In these cases, the leading order approximation gives out
\be \label{4.2}
D_0 \Psi \cong \frac{\rm d}{{\rm d} t} \Psi, ~~~
D_i \Psi \cong \frac{\partial}{\partial x^i} \Psi.
\ee
For neutron, it carries no electric charge, therefore,
equation (\ref{6.17}) is changed into
\be \label{4.3}
i \hbar \frac{\rm d}{{\rm d} t} \Psi = \left\lbrack
 \frac{1}{2m} \left (-i \hbar \nabla + m g \svec{C}^0 \right )^2
+ m g C_0^0
- \frac{g}{2} \svec{\sigma} \cdot \svec{B}^0
- \frac{i g}{2}  \svec{\sigma} \cdot \svec{E}^0
 \right\rbrack \Psi.
\ee
In the above equation, the Plank constant $\hbar$ is clearly
written out.
\\

Now, starting from this schrodinger equation,
we will discuss related quantum  effects of gravity.
From the equation (\ref{4.3}), we could see that when a neutron
is moving in gravitational field, the term $m g \svec{C}^0$
will contribute a phase factor
\be \label{4.4}
\delta \phi_1 = - \frac{m g}{\hbar} \int_{}^{r}
    \svec{C}^0 (\svec{r}')  \cdot {\rm d} \svec{r}'
\ee
Supposed that there a interference neutron beam split into two
beams at point A and recombined at point B. One beams goes along
the path $C_1$ and another goes along the path $C_2$. When two
beams recombined at point B, the phase difference between two
beams will be
\be \label{4.5}
\delta \phi_1 = - \frac{m g}{\hbar}
    \left ( \int_{C_1}^{} - \int_{C_2} \right  )
    \svec{C}^0 (\svec{r}')  \cdot {\rm d} \svec{r}'
    = - \frac{m g}{\hbar} \oint_{C}^{}
    \svec{C}^0 (\svec{r}')  \cdot {\rm d} \svec{r}',
\ee
where $C$ is the closed path formed by path $C_1$ and $C_2$.
Define
\be \label{4.6}
\Phi = \oint_{C}^{} \svec{C}^0 (\svec{r}')  \cdot {\rm d} \svec{r}'
    = - \int \! \! \! \! \int \svec{B}^0 \cdot {\rm d} \svec{\sigma},
\ee
therefore, the phase difference can be written as
\be \label{4.7}
\delta \phi_1 = - \frac{m g }{\hbar c} \Phi.
\ee
This is the gravitational Aharonov-Bohm effect.\\

The classical Newtonian gravitational potential $m g C_0^0$ will
contribution another phase factor
\be \label{4.8}
\delta \phi_2  = - \frac{m g }{\hbar}
    \int C^0_0 {\rm d} t
    = - \frac{m g}{\hbar} \int C^0_0 \frac{{\rm d} x}{v},
\ee
where $v$ is the velocity of the neutron.
For the closed path $C$ discussed above, the classical Newtonian
gravitational potential $m g C_0^0$ will cause a phase
difference
\be \label{4.9}
\delta \phi_2  = - \frac{m g }{\hbar}
    \left ( \int_{C_1} - \int_{C_2} \right )
    \frac{C^0_0}{v} {\rm d} x .
\ee
$\delta \phi_2$ is just the gravitational phase detected
in COW experiments\cite{1,2,3}. \\

The spin coupling term $- \frac{g}{2} \svec{\sigma} \cdot \svec{B}^0$
in the Schrodinger equation (\ref{4.3}) will contribute the third
phase factor
\be \label{4.10}
\delta \phi_3  =  \frac{g }{2 \hbar} \int
     \svec{\sigma} \cdot \svec{B}^0 {\rm d} t
    =  \frac{g}{2 \hbar} \int \svec{\sigma}
    \cdot \svec{B}^0 \frac{{\rm d} x}{v},
\ee
For the closed path $C$, this term will cause a phase differernce
\be \label{4.11}
\delta \phi_3  =  \frac{g }{2 \hbar}
    \left ( \int_{C_1} - \int_{C_2} \right )
    \frac{\svec{\sigma} \cdot \svec{B}^0}{v}~ {\rm d} x.
\ee
This phase originate from the coupling between spin and
gravitomagnetic field. \\

When the moving speed of neutron is very slow and the gravitational
field is weak, the magnitude of last two terms in the right-hand
side of the Schrodinger equation is much smaller than that of the
classical Newtonian gravitational potential $m g C^0_0$. So, the
above Schrodinger equation can be further simplified to
\be \label{4.12a}
i \frac{\rm d}{{\rm d} t} \Psi = \left\lbrack
- \frac{\nabla^2}{2m}
+ m g C_0^0
 \right\rbrack \Psi.
\ee
This is just the Schrodinger equation that is widely used.
For a static problem, this Schrodinger equation becomes
\be \label{4.12}
 \left\lbrack
- \frac{\nabla^2}{2m}
+ m g C_0^0
 \right\rbrack \Psi = E' \Psi.
\ee
In a local region in the earth's surface, the Newtonian gravitational
potential is a linear function of the altitude $z$, that is
\be \label{4.13}
m g C^0_0 = V_0 + m g_a z,
\ee
where $g_a$ is the gravitational acceleration and $V_0$ is
the gravitational potential at the origin. For experiments
on the earth, $V_0$ is the gravitational potential at earth's
surface. Then, the Schrodinger equation (\ref{4.12}) will
be changed into
\be \label{4.14}
 \left\lbrack - \frac{\nabla^2}{2m}
+ m g_a z
 \right\rbrack \Psi = E \Psi,
\ee
where $E=E'-V_0$. It's eigenvalue equation is\cite{22}
\be \label{4.14}
 \left\lbrack - \frac{\nabla^2}{2m}
+ m g_a z
 \right\rbrack \Psi_n = E_n \Psi_n,
\ee
where the energy eigenvalue $E_n$ is
\be \label{4.15}
E_n = \left ( \frac{\hbar ^2 m g_a^2}{2}
\right ) ^{1/3}
\left \lbrack \frac{3 \pi}{2}
\left ( n - \frac{1}{4} \right )
\right \rbrack ^{2/3}.
\ee
The gravitationally bound quantized states found recently\cite{4,5}
is just the eigenstate $\Psi_n$ of the above Schrodinger equation.
\\

When there is strong gravitomagnetic field but no electromagnetic
magnetic field, there is coupling  between spin and gravitomagnetic
field, which has the following coupling energy
\be \label{4.16}
 - \frac{g}{2} \svec{\sigma} \cdot \svec{B}^0.
\ee
When spin transition from down to up in  gravity, it will radiate
the following energy
\be \label{4.17}
\Delta E = g | \svec{B}^0 |.
\ee
Detecting such kind of radiation can directly measure the
gravitomagnetic field on the surface of the star.
\\

\section{Summary and Discussions}

In this paper, the Schrodinger equation for a Dirac particle
in gravitational field is obtained from the non-relativistic
limit of Dirac equations. In this equation, the traditional
Newtonian potential appears as the potential of Schrodinger
equation, which can lead to the formation of a bound state
when a particle is
in gravitational field. Besides, there is direct coupling
term between spin and gravitomagnetic field.
Possible quantum effects of a particle in gravitational
field, including three different kinds of phase effects,
are studied. \\

In fact, there are two kinds of quantum effects of gravity,
one is the effect of the quantized gravitational field,
another is the quantum effect of a particle in gravitational
field. In this paper, the second kind of quantum effects of
gravity is studied. These effects should be observable in some
astrophysical processes\cite{23}.
Gravitational wave and gravitational shielding effects\cite{17}
are belong to the effects of quantized gravitational
field. \\

From the study of  the elementary particle physics, we know that
the effective coupling constant of a fundamental interactions is
not a constant, it is a function of the energy scale $E$, that is
$g=g(E)$. For classical gravitational interactions, gravity is
transmitted by virtual graviton. In this case, the effective coupling
constant is the infrared limit of $g(E)$, that is $g(0)$. For
quantum gravitational interactions, when the energy scale
is much larger than zero, the effective coupling constant should
be much larger than $g(0)$, and the quantum effects of the quantized
gravitational field can be much larger than traditionally expected.
Generally speaking, it is not correct to
use Newtonian gravitational coupling constant to estimate the order
of the magnitude of the quantum gravitational interactions.
For the present situation, it is important to calculate the
$\beta$-function of gravitational interactions based on quantum
gauge theory of gravity.
\\

\end{document}